\documentclass[12pt]{article}
\makeatletter
\@addtoreset{equation}{section}

\newcommand{\p}{\partial}

\begin{document}

\begin{titlepage}
\vfill
\begin{flushright}
{\normalsize DAMTP-2002-101}\\
{\normalsize UT-Komaba/02-07}\\
{\normalsize KIAS-P02051}\\
{\normalsize\tt hep-th/0209034}

\end{flushright}

\vfill
\begin{center}
{\Large\bf
Tachyon Condensates, Carrollian Contraction of Lorentz Group,
and Fundamental Strings
}

\vskip 0.3in

{
\large
Gary Gibbons$^a$\footnote{\tt G.W.Gibbons@damtp.cam.ac.uk},  
Koji Hashimoto$^b$\footnote{\tt koji@hep1.c.u-tokyo.ac.jp},
and 
Piljin  Yi$^c$\footnote{{\tt piljin@kias.re.kr}}
}

\vskip 0.15in

${}^a$ {\it D.A.M.T.P., Cambridge University, Wilberforce Road,} \\
{\it Cambridge CB3 0WA,  U.K.}\\[3pt]
${}^b$ {\it Institute of Physics, University of Tokyo, Komaba}\\
{\it Tokyo 153-8902, Japan}\\[3pt]
${}^c$ {\it School of Physics, Korea Institute for Advanced Study,} \\
{\it 207-43, Cheongryangri-Dong, Dongdaemun-Gu, Seoul 130-012, Korea}
\\[0.3in]

{\normalsize September, 2002}

\end{center}

\vfill

\begin{abstract}
\normalsize\noindent 
We study the  rolling tachyon condensate in the presence of a gauge
 field. The generic vacuum admits both a rolling tachyon, $\dot T$, and
 a uniform electric  field, ${\vec E} $, which together affect the
 effective metric governing the  fluctuations of open string modes. If
 one suppresses the gauge field altogether, the light-cone collapses
 completely. This is the Carrollian limit, with vanishing speed of light
 and no possible propagation of signals. In the presence of a gauge
 field,
 however, the lightcone is squeezed to the shape of a fan,  allowing
 propagation of signals along the direction of $\pm \vec E$ at speed
$|\vec E| \le 1$. This shows that there are perturbative degrees of freedom
 propagating along electric flux lines. Such causal behavior appears to
 be a very general feature of tachyon effective Lagrangian with runway
 potentials. We speculate on how this may be connected to appearance
 of fundamental strings.
\end{abstract}

\vfill

\end{titlepage}
\setcounter{footnote}{0}

\pagebreak
\renewcommand{\thepage}{\arabic{page}}
\pagebreak

\setcounter{equation}{0}

\section{Effective Field Theory of Open String Tachyon Condensation}

The decay of unstable D-branes \cite{DaD} has served as an important
laboratory  
in understanding the  off-shell physics of open string theory. One of
the  more
surprising aspect of this story is how powerful the tree-level
theory is in reproducing the desired feature of the process \cite{emil}. 
For example, the 
so-called Boundary String Field Theory (BSFT) \cite{BSFT, BSFTo} 
has reproduced the exact height of the tachyon potential as 
well as exact tension of some of the lower dimensional D-branes as 
solitons \cite{GS, kutasov}. 
This sort of result seems pretty surprising, since the effective
field theory one employs is really only  justified at the initial stage 
of the process near the top of the tachyon potential. We might expect
 that the effective field theory description of
the process would be useless near the bottom of the potential, 
where open string degrees of freedom, or field content thereof, 
does not make much sense. 

One hint that the use of classical effective field theory is not perhaps 
completely unjustified, is the fact that the field theory seems
to lack most of the  perturbative degrees of freedom. A toy example
of this was given recently by A. Sen \cite{rolling} where he considers a 
Lagrangian of type
\begin{equation}
-V(T)\sqrt{-\det(\eta+\partial T\partial T)}
=-V(T)\sqrt{1+(\partial T)^2}.
\end{equation}
This form of action was first proposed by Garousi and also by 
Bergsheoff et al.\cite{bergshoeff} and 
explored previously in Refs.\ \cite{kluson,fluid}.
With the runaway form of potential $V\sim e^{-T}$ or $e^{-T^2}$,
which is also natural in the BSFT type action,
Sen notes that the perturbative, planewave-like solutions are completely
absent as the system approaches the vacuum, $T=\infty$. At the bottom
of the potential, the D-brane should have disappeared, and thus no
open string degrees of freedom should be possible. Whatever the 
weaknesses  of
the effective field theory there might be, it nevertheless  captures
some of 
the crudest aspect of the tachyon condensation. Much of this depends
crucially on the unusual nature of the vacuum, which requires the
tachyon field to evolve in time as $\dot T=1$.

On the other hand, a different sort of behavior had been found in 
Ref.~\cite{fluid}
where the authors considered an effective field theory with Lagrangian
\cite{potential}
\begin{equation}
-V(T)\sqrt{-\det (\eta+F)},
\end{equation}
where one takes $T$ as non-dynamical and instead concentrate on 
the worldvolume gauge field. With this system, the counterpart of
$\dot T=1$ above is that the electric field goes critical $|\vec E|=1$.

The dynamics of this system was found to
be that of string-like fluid, consisting of remnant of electric
flux lines, rendered pressureless in the limit of $V\rightarrow 0$.
Even though the fluid is pressureless, similar to the tachyon matter,
one crucial difference is that the basic object is one-dimensional
and small fluctuation of such one-dimensional object is still
allowed. Thus, the dynamics is not completely trivial and 
perturbative degrees of freedom survive. In fact, an
intriguing aspect of this dynamics was that it contains a special
subclass of solutions that resembles relativistic Nambu-Goto string
\cite{fluid,followup},
allowing the  wishful thought that fundamental, closed string might 
emerge naturally from some effective field theory of tachyon 
condensation. More details about string fluids can be found in
\cite{Gibbons2}. Over the years, the issue of closed strings has proved
to be one of most difficult question to answer in the open string
tachyon condensation \cite{fluid,yi,conf,shenker}.

In this paper, we would like to explore a general class of effective
field theories  of the form,
\begin{equation}
-V(T)\sqrt {-\det (\eta+F)}\;{\cal F}(z),
\end{equation}
with 
\begin{equation}
z\equiv ((\eta+F)^{-1})^{\mu\nu}\;\partial_\mu T\partial_\nu T.
\end{equation}
With ${\cal F}(z)=\sqrt{1+z}$, this includes the above two cases
as special limits, while with yet another choice of ${\cal F}$, the
effective action is exactly the one found in the  BSFT approach. In this
general class of action with both gauge field and tachyon field
present, the vacuum condition is modified to $\vec E^2+\dot T^2=1$,
which interpolates between the two extreme cases above. The
main purpose of this paper is to explore this class of 
effective field theory and ask what kind of classical dynamics
it possesses. 

The notion of the effective metric proves to be quite
useful in this endeavor. In nonlinear theories, the causal structure of 
signal propagation does not coincide with that of the background 
spacetime (or worldvolume). Rather one must find the effective 
metric by expanding the action or the field equation. A typical 
behavior we find is that such an effective metric tends to collapse 
to a singular form, and restricts possible propagation of signal 
to some extreme forms. In the purely tachyonic case above, this
collapse is isotropic and effectively sets "the speed of light"
to zero, as we will see shortly. This singular limit, which
can be thought of as a limit of the Lorentzian spacetime, opposite
to the Galilean limit, is known as Carrollian. This collapse
of the causal structure is precisely what is responsible for
disappearance of perturbative degrees of freedom.

With gauge field turned on, however, the story changes qualitatively.
While the lightcone of small fluctuation still collapses, it does
leave some directions open for propagation of signal; Instead of
the isotropic collapse of the effective metric, one finds that
the causal structure goes Carrollian in all spatial directions except 
one, chosen by the background electric field, $\vec E$. Along this
special direction only, small fluctuations propagate at speed 
$|\vec E|$. 
The bulk of this paper will be devoted to derivation of this fact, and
also to its implication to tachyon condensation.

In section 2, we briefly recall the Carroll limit as a limit of 
vanishing speed of light. In section 3, we derive the effective
metric of the purely tachyonic system, and demonstrates that the
lightcone collapses completely and the effective metric is 
Carrollian. In section 4, we explore the effective
field theory with gauge fields included, and demonstrate that
some perturbative degrees of freedom remain and propagates along
one special direction. In section 5, we employ the Hamiltonian formalism  
and consider the dynamics of small fluctuation as well as that of 
electric flux lines. For the sake of simplicity we restrict to the
case of the action of Ref.~\cite{bergshoeff} in this section.
In particular, we study how the dynamics of the fundamental string 
flux lines are affected by the singular causal structure.
In section 6, we conclude with summary and directions for future
research.

\section{Carroll Group}

The Lorentz group $SO(3,1)$ and its extension by spacetime
translations, the  Poincar\'e group $E(3,1)$, are central to our
understanding of  issues such as  causality  in quantum field
theory. In fact there is not a single Lorentz group but one for
each value of the absolute velocity $c$. Of course they are all
isomorphic except in the In\"on\"u-Wigner limiting cases when 
$c \uparrow \infty$ or $\downarrow 0$. The former case corresponds to
the Galileo group when we have instantaneous propagation and
action at a distance with field satisfying elliptic partial
differential equations, the latter, which is less well known, is
called the Carroll group \cite{Leblond, Sen} and corresponds to
the case of no propagation at all. Fields at each spatial point
evolve independently and are typically governed by ordinary
differential equations with respect to just the time variable. For
that reason, this case often arises as the symmetry group of an
approximation scheme in which spatial derivatives are ignored
compared with time derivatives. Such approximation schemes are
sometimes called ``velocity dominated''.

Geometrically the Galileo group arises when the future light cone
flattens out to become a spacelike hyperplane. The Carroll group
arises when it collapse down to a timelike half line. In the
Galilean case only the contravariant metric tensor has a well
defined limit as $c \uparrow \infty$:

\begin{equation} 
\eta ^{\mu \nu} \rightarrow {\rm diag} (0,1,1,\dots ,1) 
\end{equation}
and the limiting spacetime structure is called a Newton-Cartan
spacetime. In the case of the Carrollian limit it is the covariant
metric tensor which survives
\begin{equation} 
\eta_{\mu \nu } \rightarrow {\rm diag} (0,1,1,\dots, 1),
\end{equation}
and one has a Carrollian spacetime.

A striking feature of the
Carroll group is that the contracted Lie algebra contains a
Heisenberg sub-algebra in which energy $P_0$ appears as a central
charge. In other words, if $P_i$ generate the spatial translations
and $B_i$ the boosts, then the only non-vanishing commutator among
these  7 generators is
\begin{equation} 
\Bigl [ P_i, B_j\Bigr ]= \delta_{ij} P_0.  
\end{equation}
Of course the spatial rotations $J_{ij}$ act in the usual way. In
four spacetime dimensions,  there are two Casimirs, the energy
$P_0$ and $(P_0  {\vec  J }+  {\vec  P} \times {\vec B} ) ^2$.

Although it emerges naturally in any classification of possible
kinematic groups \cite{Bacry, Nuyts}, as of now,  the Carroll
group has played a rather minor role in physics. In this paper we
explore the possibility that it enters in an essential way into
the phenomenon of tachyon condensates in string theory. If the
latter  turn out to have a cosmological role \cite{Gibbons1,cos}, then
the Carroll group will certainly come into its own sometime in the
future.

To illustrate  how a field theory may behave in a Carrollian limit,
let us consider a Maxwell field in 3+1 dimensions.
The standard Lorentz-invariant Maxwell equations are, after inserting
the velocity of light $c$, 
\begin{equation}
\frac{\partial {\vec E}}{\partial t } = c^2   {\rm curl} \thinspace 
{\vec B},
\qquad  \frac{\partial  {\vec B}}{\partial t} = - {\rm curl}
\thinspace {\vec E},  
\end{equation} 
and
\begin{equation}
{\rm div} \thinspace {\vec E}=0, \qquad {\rm div} \thinspace {\vec B}=0.
\end{equation} 
If we now take the limit $c \downarrow 0$, we get
\begin{equation}
\frac{\partial {\vec E} }{ \partial t } = 0
\qquad  \frac{\partial  {\vec B} }{ \partial t} = - {\rm curl}
\thinspace {\vec E},  
\end{equation}
Thus in the Carrollian limit the electric field $\vec E$ is time
independent
\begin{equation}
{\vec E({\vec x}, t) } = {\vec E} ({\vec x}, 0). 
\end{equation}
By contrast, the magnetic field $ {\vec B} $ evolves linearly in time:
\begin{equation}
{\vec B}({\vec x}, t)= {\vec B} ( {\vec x}, 0) - t \thinspace {\rm curl} 
\thinspace {\vec
E} ({\vec x}, 0).
\end{equation} 
In a later section, we will encounter a similarly degenerate form of
gauge field theory, which differs from this example in two important 
aspects. One is that it derives from the nonlinear Born-Infeld action,
and the other is that its causal behavior will be slightly different
from this Carrollian limit, in that along some special direction, set 
by the background "electric field" $\vec E$, the effective speed of
light remains finite. See section 4 and 5 for more detail.

\section{Effective Metric for Tachyon Fluctuation}

\subsection{Open and Closed String metrics}

In string theory there are closed strings states and open string
states. The former include the graviton and their propagation is
governed by the Einstein metric $g_{\mu \nu}$ or the conformally
rescaled string metric. Both have the same light cone and both
define the same local Lorentz group. It seems that all closed
string states  couple to the same metric and hence have the same
causal behavior. In other words  a form of the usual equivalence
principle holds for closed string states. Because closed string
states  couple to the dilaton the full Einstein Equivalence
principle will not hold if the dilaton remains massless.

By contrast, open string states are governed by a different
metric, the open string metric ${ G^{\rm open}}  _{\mu \nu}$. In
the case of D-branes described by a Dirac-Born-Infeld action for
example the open string metric differs from the induced string
metric $_\star g_{\mu \nu} $  by terms which depend upon the
gauge-invariant combination 
${\cal F} = F_{\mu \nu} - {}_\star B_{\mu \nu}$, 
where $F_{\mu \nu}$ is the Born-Infeld field strength  and
$_\star B_{\mu \nu}$ is the  Neveu-Schwarz potential pulled back
to the brane world volume. It seems that all open string states
propagating along the brane also satisfy a form of the equivalence
principle, their causal properties are all governed by the light
cones of the open string metric $G_{\mu \nu}$. Thus in addition to
the closed string local Lorentz group there is an open string
local Lorentz group. How are the two light cones related? One
finds that in the Born-Infeld case, although they may
coincide, the open string light cones never lie outside the closed
string light cones \cite{Herdeiro, Gibbons2, Gibbons3}. The same
holds true for the M-theory 5-brane \cite{West}.

The comparisons made above are between signals, both of which
propagate in the brane. A different question, which arises in the
AdS/CFT correspondence is the time delay suffered by signals
travelling through the bulk compared with those restricted to the
boundary. It seems that general principles dictate that the signal
travelling in the bulk should indeed always be delayed rather than
advanced \cite{Wald, Page}. The distinction between bulk and brane
disappears in the case of space-filling branes and in this paper
we shall ignore it.

\subsection{Effective Metric for Tachyon Fields}

Now if considers a classical  tachyon field $T(x)$ there is the
possibility of yet another metric arising, or perhaps one
should think of the tachyon as modifying the open string metric.
The general Tachyon Lagrangian density $\cal L$ in the absence of
Born-Infeld etc is of the form

\begin{equation} 
{\cal L} =\sqrt{-g} \thinspace  L(T, y),  
\end{equation}
where
\begin{equation} 
y= g^{\mu \nu}
\partial_\mu  T \partial_\nu T, 
\end{equation} 
and as usual $g = {\rm det}\thinspace g_{\mu \nu}$.

Thus for example Sen chooses \cite{rolling}
\begin{equation} 
{\cal L}= -\sqrt{-g} \thinspace
V(T) \sqrt{1+y} = -V(T) \sqrt{-\det \bigl(  g_{\mu \nu}
+\partial_\mu T
\partial_\nu T \bigr ) },  
\label{sencho}
\end{equation} 
where $V(T)$ is the tachyon potential.
However other Lagrangians have been considered and in any case the
classical Lagrangian is scheme dependent. For example the BSFT
action for a non-BPS D-brane in superstring theory 
is, in the absence of the Born-Infeld field, \cite{kutasov}
\begin{equation}
{\cal L} = - \sqrt {-g} \thinspace  e^{ - \frac{1}{4} T^2 }
{\cal F}(y), 
\end{equation} 
where 
\begin{equation} 
{\cal F}(y) = \frac{y 4^y \Gamma (y) ^2}{2 \Gamma (2 y) }.
\label{bsftf}
\end{equation}
In the Lagrangians we 
omit the overall tension $T_p$ of the non-BPS D$p$-brane,
and we work in the unit $\alpha'=2$.

Therefore in the sequel we shall consider the general case. The
energy momentum tensor is given by 
\begin{equation} 
T^{\mu \nu} = L g^{\mu \nu} -2 L_y \nabla ^\mu T 
\thinspace \nabla ^\nu T, 
\end{equation} 
where 
$\nabla ^\mu T = g^{\mu \nu }\partial_\nu T$ and the subscript $y$
denotes partial differentiation with respect to $y$. If  $T$ is
just a function of time, $y=-\dot T^2 $ and as it rolls, the
tachyon energy density $\rho$ and pressure $P$ are given by 
\begin{equation}
\rho = 2yL_y -L, \qquad P= L. 
\end{equation}

The weak energy condition, that is $T^{\mu \nu} p_\mu p_\nu \ge 0$
for all co-vectors $p_\mu$ which are causal with respect to the
metric $g_{\mu \nu}$  will hold if  $ 2y L_y-L \ge 0$. The
Dominant Energy Condition, that is $T^{\mu \nu} p_\mu$ is future
directed timelike or null with respect to the metric $g_{\mu \nu}$
for all future directed timelike or lightlike covectors $p_\mu$
will hold if $ |L| \le 2yL_y -L$. The Dominant Energy condition
guarantees a degree of causal behavior in that, as originally
proved by Hawking \cite{Hawking} (see also Refs.\
\cite{Gibbons2, Carter} ),  
if the tachyon energy density vanishes outside some
compact set, then it vanishes outside the future of that set. In
other words an advancing front of  energy density advances into
vacuum at a speed no faster than that of light. A rather different
question is how fluctuations around a  tachyon background travel.

To answer that,  recall that the tachyon equation of motion may be
written as 
\begin{equation} 
\bigl ( G^{-1} \bigr )  ^{\mu \nu} \nabla _\mu
\nabla _\nu T = \frac{L_T}{2 L_y} , 
\end{equation}
where \begin{equation} 
\bigl (G^{-1} \bigr ) ^{\mu \nu} = g^{\mu \nu} + 
\left( \frac{2 L_{yy}}{L_y} \right)  \nabla ^\mu T \thinspace \nabla
^ \nu T 
\label{inverse}
\end{equation} 
is the inverse of the metric
\begin{equation} 
G_{\mu\nu} = g_{\mu \nu} - 
\left( \frac{2 L_{yy}}{L_y + 2 y L_{yy}}\right)
\partial _\mu \thinspace T \partial _\nu T .  
\label{inversem}
\end{equation}

Evidently the characteristics of the tachyon field are given by
the light cones of the metric $G_{\mu \nu}$. In other words the
speed of small fluctuations around a tachyon field background is
governed by the ``tachyon metric'' $G_{\mu \nu}$.   By considering
a co-vector $p_\mu$ which is lightlike with respect to the metric
$g_{\mu \nu}$, that is such that $g^{\mu \nu} p_\mu p_\nu =0$, one
discovers that
\begin{equation} \bigl ( G^{-1}\bigr ) ^{\mu
\nu} p_\mu p_\nu=  \left(\frac{2L_{yy}}{L_y}\right)  \bigl(p_\mu
\nabla ^\mu T \bigr ) ^2 . 
\end{equation} 
It follows that if $2 L_{yy} / L_y$ 
is negative then the propagation of fluctuations of the
tachyon field will be slower than light. This is actually the case
for Sen's action (\ref{sencho}) in the region $-1 < y$.

The energy momentum tensor and metrics for the BSFT case is
somewhat complicated. It is known to satisfy both Weak Energy
condition and the Dominant  Energy condition. One may also check
that the light cones of the metric  $G_{\mu \nu}$ never lie
outside the light cone of the metric $g_{\mu \nu}$. 
In fact, the quantity $2L_{yy}/L_y$ is always negative for 
$-1< y \leq 0 $ which is the region relevant for the rolling tachyon
in the BSFT case \cite{Terashima}.

\subsection{The Carrollian Behavior of Tachyonic
Fluctuations in Rolling Tachyon Background}

In the case of Sen's choice of tachyon action, 
with ${\cal F}(y)=\sqrt{1+y},$ 
one finds 
\begin{equation}
G_{\mu \nu} = g_{\mu \nu}  + \partial _\mu T \thinspace \partial
_\nu T, 
\end{equation} 
and 
\begin{equation} 
\bigl ( G^{-1} \bigr ) ^{\mu \nu} = g^{\mu
\nu} -\frac{1}{1+y} \nabla^\mu T \thinspace \nabla^\nu T.
 \end{equation}
Moreover the energy momentum tensor takes the
strikingly simple form 
\begin{equation} 
T^{\mu \nu} = -V(T) \sqrt {1+y} \bigl
( G^{-1} \bigr )  ^{\mu \nu}= L \bigl ( G^{-1}  \bigr ) ^{\mu \nu}
. \end{equation}
In this simplest model of Sen,
the tachyon potential $V(T)$
has a minimum at infinity, and as the tachyon rolls towards it,
one finds $T \rightarrow \infty$ and $|\dot T| \rightarrow 1$. 
As
this happens we find that 
\begin{equation} 
G_{\mu \nu} \rightarrow {\rm diag }
(0,1,1, \dots, 1 ). 
\end{equation} 
Thus in the limit the metric degenerates,
the tachyon light cone collapses onto a timelike half line and
the tachyon fields at different spatial points are decoupled. No
propagation of tachyon fluctuations can take place. 
This phenomenon lies at the heart of recent observations that
no plane wave is possible for the tachyonic field near the bottom of the 
potential $T=\infty$.

In the case of the BSFT action one may use the fact that
near $y=-1$ one has \cite{Terashima}
\begin{equation}
{\cal F}(y) \sim - \frac{1}{2(y+1)},
\label{limitt}
\end{equation}
to show that as $|\dot T| \rightarrow 1$, 
$ G_{\mu \nu} \rightarrow {\rm diag} ( 0, 1, \dots ,1)$.
Thus we see that the 
BSFT metric shares with the tachyon metric $G_{\mu \nu} $ the property
that it becomes Carrollian near the tachyon vacuum. In general, it is
obvious from the expression (\ref{inversem}) that, if $L_{yy}$ is more
divergent than $L_y$ around the rolling tachyon phase $y\sim -1$, the
effective metric $G_{\mu\nu}$ becomes Carrollian.

\section{Born-Infeld Gauge Fields}

If the Carrollian metric governs all open string fluctuations then 
these also will cease to propagate. The main objective of this section
is to show that this is not the case, once we consider gauge fields.
The gauge field enters the action through a Born-Infeld piece, and one 
obvious generalization of Sen's model is\footnote{Rolling tachyon
coupled to gauge fields has been investigated also in Refs.\
\cite{Ishida, Muk}. } 
\begin{eqnarray}
{\cal L}=  - V(T) \sqrt{-  \det (g + F +\p T\p T)}.
 \label{tbi}
\end{eqnarray}
In fact, this form of action had been proposed in Ref.\ 
\cite{bergshoeff} and
studied in much detail in Ref.\ \cite{fluid}. To allow uniform treatment
of this action and the BSFT action, we rewrite this action as
\begin{equation}
{\cal L} =  - V(T) \sqrt{-  \det (g + F) } {\cal F}(z), \label{gauge}
\end{equation} 
with ${\cal F}=\sqrt{1+z}$, where 
\begin{equation} z=\left((g+F)^{-1}\right)^{(\mu \nu)}
\partial _\mu T\partial _\nu  T .
\end{equation}
Note that only the symmetric part of $(g+F)^{-1}$ enters the 
expression of kinetic term for $T$.\footnote{ 
This metric $\widetilde g$ is what one usually
calls ``open string co-metric,'' for instance, in noncommutative
setting. In the current context, with $\dot T$ nonzero, this metric is 
no longer ``open string metric''.}  Then, the BSFT action can be
written in the same form with a different ${\cal F}$, as given in
(\ref{bsftf}).

The main difference one encounters when the gauge field in included
is that vacuum structure changes substantially. It was observed in 
Ref.~\cite{fluid} that, for Sen's effective action
the determinant part 
\begin{equation}
\sqrt{-  \det (g + F +\p T\p T)} 
\end{equation}
itself should vanish as $V\rightarrow 0$. This is really why 
$\dot T = 1$ in a purely tachyonic system, while in Ref.~\cite{fluid}, 
the vacuum considered is such that $\dot T=0$ and $E=1$ instead.
In is clear that these two extreme case are connected by a continuous 
family of vacua
\begin{equation}
\dot T^2+E^2 = 1. \label{1}
\end{equation}
As will be shown shortly, this holds not only for the above
form of the action, but also for the effective action derived from BSFT
method.\footnote{The tachyon effective action proposed by 
Lambert and Sachs \cite{LS} does not possess this property. } 

In this section, we will investigate how fluctuation of the 
tachyon field and that of the gauge field behave in generic 
vacuum with both $\dot T$ and $E$ nontrivial. A single gauge
field in $p+1$ dimensions possesses $p-1$ propagating degrees
of freedom, due to the gauge invariance, while $T$ would give,
at least in ordinary circumstances, one degree of freedom.
In a nutshell, the effective metric for these $(p-1)+1=p$
fluctuation is no longer Carrollian but given by 
\begin{equation}
G_{\mu \nu} \rightarrow \lim_{\epsilon\rightarrow 0} {\rm diag }
\left(-\epsilon,\frac{\epsilon}{E^2},1, \dots, 1 \right),
\label{difca}
\end{equation} 
as the tachyon condensation proceeds. This causal structure is different
from the Carrollian limit, since along one spatial direction (chosen by
the direction of the background electric field $E$) gauge fluctuations
may propagate. Interestingly the speed at which the signal propagates 
is precisely $E$.\footnote{ One might be mislead to think that this
actually means gauge fluctuation is absent, since  in 1+1 dimensions
gauge degrees of freedom are absent. We emphasize that this is not the
case here. While propagation may be allowed only along a specific
direction, the gauge field still lives on entire worldvolume of the
brane. Nontrivial degrees of freedom comes from transverse distributions
of electric flux lines.}


\subsection{Background Equations of Motion}

First, let us show that the constant electric field is 
allowed in the rolling tachyon phase, 
as a classical solution of the tachyon system coupled to gauge field. 
We shall use the following general action which includes both 
Sen's action and the BSFT action, 
\begin{eqnarray}
  {\cal L} = -V(T) \sqrt{-\det (\eta + F)} \;
{\cal F}(z).
\label{geneact}
\end{eqnarray}
Recall that Sen's action is given by ${\cal F} (z) = \sqrt{1+z}$,
while this ${\cal F}(z)$ for the BSFT action is a complicated function 
(\ref{bsftf}) whose singular behavior around $z\sim -1$ is given by
(\ref{limitt}).
Now we turn on the electric field only along $x^1$, and assume that
all the fields are homogeneous: the electric field and the tachyon 
depend only on time. Then the Lagrangian reduces
to
\begin{eqnarray}
  {\cal L} = -V(T) \sqrt{1-E^2} {\cal F}
\left( \frac{-\dot{T}^2}{1-E^2}\right).
\end{eqnarray}
The equations of motion are given by
\begin{eqnarray}
 \dot{\pi} =0, \quad
\dot{P} + V' \sqrt{1-E^2} {\cal F} =0,
\label{eoms}
\end{eqnarray}
in which we have defined the conjugate momenta as
\begin{eqnarray}
  \pi \equiv \frac{\delta{\cal L}}{\delta E}
=V \frac{E}{\sqrt{1-E^2}} 
\left({\cal F} + \frac{2\dot{T}^2}{1-E^2} {\cal F}'
\right),
\quad 
P \equiv 
\frac{\delta{\cal L}}{\delta \dot{T}}=
2 V \frac{\dot{T}}{\sqrt{1-E^2}} {\cal F}'.
\label{P}
\end{eqnarray}
Note that $\pi$ is the electric induction which is often  denoted by
$ D$. The first equation in (\ref{eoms}) shows 
that the electric induction or flux $\pi$ is a conserved quantity.
Another conserved quantity is the energy Hamiltonian which 
is given by
\begin{eqnarray}
  {\cal H} = \pi E + P \dot{T} - {\cal L}
= V \frac{1}{\sqrt{1-E^2}} 
\left({\cal F}+ \frac{2\dot{T}^2}{1-E^2}
{\cal F}'\right).
\label{H}
\end{eqnarray}
{}From these expression, one observes an interesting relation
\begin{eqnarray}
  \pi = E {\cal H}. 
\label{rela}
\end{eqnarray}
Since both 
$\pi$ and ${\cal H}$ are conserved quantities, the constant $E$ is
a consistent solution of the system.

In particular, If we choose the initial
condition for the rolling as
\begin{eqnarray}
  T\bigm|_{t=0}=T_0, \quad \dot{T}\bigm|_{t=0}=0,
\end{eqnarray}
where $T_0$ is a positive constant, 
then the conserved energy can be evaluated at $t=0$ as
\begin{eqnarray}
  {\cal H} = \frac{V(T_0){\cal F}(0)}{\sqrt{1-E^2}}.
\end{eqnarray}
Therefore with the above relation (\ref{rela})
we have
\begin{eqnarray}
  E = \frac{\pi}{\sqrt{V(T_0)^2 {\cal F}(0)^2+\pi^2}}.
\end{eqnarray}
This means that the infinite condensation of the electric flux
$\pi=\infty$ is equivalent to  the limit $E=1$. 

As the tachyon rolls down the potential hill $V(T)$, it approaches the
true vacuum at which the potential vanishes, $V(T\!=\!\infty)=0$.
Therefore, it is obvious from the expression for $\pi$ 
(\ref{P}) and ${\cal H}$ (\ref{H}) that the combination 
\begin{eqnarray}
\left({\cal F}+ \frac{2\dot{T}^2}{1-E^2}
{\cal F}'\right)
\label{combif}
\end{eqnarray}
should be divergent, to maintain the constancy of  $\pi$ and ${\cal H}$.
For both Sen's action and the BSFT action, this divergence is
achieved when the argument $z$ of the function ${\cal F}$ approaches
$-1$ as in the case of the rolling tachyon without the electric field. 
The equation $z=-1$ gives the rolling tachyon solution with a constant
electric field, 
\begin{eqnarray}
  \dot{T}^2+E^2 = 1.
\label{conse}
\end{eqnarray}
It follows from this expression that the critical electric field $E=1$
is a special case: in this limit the flux $\pi$ diverges and the physics 
may be dominated by the fundamental string. The tachyon does not roll,
and the situation is similar to that of the supertubes \cite{supertube}.  
In the supertube configurations, tachyons caused by a brane-antibrane
configuration become massless and the whole system is
stable due to the critical limit of the electric field. 

\subsection{Unidirectional Propagation of Small Fluctuations}
\label{sena}

After having gone through previous section, one might expect that the
only change as far as the tachyon fluctuation goes,  is that 
$g^{\mu\nu}$ is replaced by $(\widetilde g)^{-1}$ which is the
symmetric part of $(g+F)^{-1}$. For Sen's action, this observation 
would give
to the effective metric
\begin{eqnarray}
G_{\mu \nu} = \widetilde g_{\mu \nu}  + \partial _\mu T 
\thinspace \partial_\nu T \rightarrow 
{\rm diag } (0,1-E^2,1, \dots, 1 ).  
\end{eqnarray}
which is again Carrollian for $E\neq 0$.
However, this naive expectation is incorrect: the gauge fluctuations and 
the tachyon fluctuation do not in general separate nicely. Rather they
are mixed together so that we have to consider them simultaneously.
In the following, we show that the tachyon and gauge fluctuations are
mixed and all $(D-1)$ degrees of freedom propagate, along a
particular direction determined by the background electric field.


First in order to 
illustrate the point, let us analyze Sen's action. 
In this action, the kinetic term for  the tachyon $T$ appears in the
square root as if it were one of the transverse scalar field. The
transverse scalar field in the Dirac-Born-Infeld action can be treated
as one of the additional gauge field \cite{BIparticle} 
through the T-duality along the
transverse direction. Then this means that we can regard $T$ as a
transverse scalar field except that it experiences the 
potential $V(T)$. However, this potential term is irrelevant for our
analysis, since we need only the causal behavior which is encoded in the
kinetic term. The only information which we need from the potential term
is that we are going to the rolling tachyon limit $z\rightarrow -1$. 
Therefore, for our purpose, it is sufficient to consider the
Born-Infeld equations of motion in one higher dimensions $p+2$ :
\begin{eqnarray}
\left(
\frac{1}{\eta + F}
\right)^{(\mu\nu)} \p_\mu F_{\nu\rho}=0.
\label{dbieq}
\end{eqnarray}
Here note that the gauge field strength is that in $p+2$ dimensions. The
indices are $\mu, \nu,\rho = 0,1,\cdots,p,T$, and 
\begin{eqnarray}
 F_{\mu T} \equiv \p_\mu T.
\end{eqnarray}
We use the above equations of motion with the understanding that 
$\p_T=0$. 

Let us define a set of 
new worldvolume coordinates which turn out to be useful,  
\begin{eqnarray}
 \left(
\begin{array}{c}
\widetilde{x}^T \\
\widetilde{x}^1
\end{array}\right)
\equiv 
\frac{1}{\dot{T}^2 + E^2}
\left(
\begin{array}{cc}
\dot{T} & E 
\\ -E & \dot{T}
\end{array}
\right)
 \left(
\begin{array}{c}
x^T \\
x^1
\end{array}\right).
\label{trot}
\end{eqnarray}
This redefinition is the ``target space rotation'' which was
investigated in Ref.\ \cite{Hashimoto}. The origin of this rotation is
as follows. First, let us take a different T-duality, not along the
transverse direction but along $x^1$. Then the world volume is $p$
dimensional, and we have two transverse scalar fields, $T$ and $A_1$. 
In this T-dualized  language, the rolling tachyon limit 
\begin{eqnarray}
 \dot{T} \rightarrow \sqrt{1-E^2}
\end{eqnarray}
can be understood as a light-like limit of the worldvolume, i.e.\ the 
$(p\!-\!1)$-brane is travelling with the speed of light along the
direction  
$(x^1, x^T)= (E, \sqrt{1-E^2})$. Our target space rotation 
(\ref{trot}) is just the rotation to orient the brane motion into the
direction along $\widetilde{x}^T$.

It is known that when the
worldvolume is lightlike then the induced metric on the brane is
Carrollian. 
In our case, we have to take a T-duality back to have $p+1$
dimensional world volume, thus the question is not so simple. Actually,
we will find that the dynamics in the rolling tachyon limit is not
Carrollian: a part of the gauge fields become dynamical and
can propagate along $x^1$, with the effective metric 
(\ref{difca}).\footnote{A simple reason may be that 
the last T-duality is not orthogonal to the direction of the
motion of $(p\!-\!1)$-brane. It induces the electric field back 
\cite{Bachas} and
non-trivial dependence on the T-dualized direction ($x^1$). }

By the target space rotation, the gauge field strength is transformed
as usual, 
\begin{eqnarray}
  \widetilde{F}_{\tilde{\mu}\tilde{\nu}} = 
\frac{\p x^\rho}{\p \widetilde{x}^\mu}
\frac{\p x^\sigma}{\p \widetilde{x}^\nu} F_{\rho\sigma},
\end{eqnarray}
therefore the tensor appearing in (\ref{dbieq}) described by the new
coordinates becomes 
\begin{eqnarray}
\left(
\frac{1}{\eta + \widetilde{F}}
\right)_{\rm sym}
&=& \left(
\begin{array}{ccc}
-1 & 0 \cdots 0 & E^2 + \dot{T}^2 \\
0 & \cdots & 0 \\
-E^2 -\dot{T}^2 & 0 \cdots 0 & 1
\end{array}
\right)^{-1}_{\rm sym}
\nonumber \\[2pt]
&=& \mbox{diag}
\left(
\frac{-1}{1-E^2-\dot{T}^2}, 1, \cdots,1, 
\frac{1}{1-E^2-\dot{T}^2}
\right).
\end{eqnarray}
In the limit $\dot{T} \rightarrow \sqrt{1-E^2}$ the equations
of motion (\ref{dbieq}) reduces to the two dimensional one, 
\begin{eqnarray}
 \p_0 \widetilde{F}_{0 \widetilde{T}} - \p_{\widetilde{T}}
\widetilde{F}_{\widetilde{T}\widetilde{\mu}}=0.
\label{origin}
\end{eqnarray}
Since we are studying the fluctuation around the rolling tachyon
background, we expand the fields as 
\begin{eqnarray}
 T = T_0 + t, \quad 
  F_{\mu\nu} = E 
(\delta_{\mu 0} \delta_{\nu 1} - \delta_{\mu 1} \delta_{\nu 0})
+ f_{\mu\nu},
\end{eqnarray}
where $f_{\mu\nu}= \p_\mu a_\nu - \p_\nu a_\mu$, and $t$, $a_\mu$ are
the fluctuations. 
Then the causal structure for the fluctuation can be extracted from 
(\ref{origin}) if one replace $\widetilde{F}$ by its fluctuation and 
(\ref{trot}) by its background value, since we need only the term in
which the fluctuations receives two derivatives. We shall work in the
gauge $a_0=0$.

For $\widetilde{\mu}=\widetilde{T}$, the equation (\ref{origin}) is
reduced to
\begin{eqnarray}
  (\p_0)^2
\left[ E a_1 + \sqrt{1-E^2}\; t\right]=0.
\label{hatt}
\end{eqnarray}
Hence if we define a linear combination
\begin{eqnarray}
 \hat{t} \equiv \sqrt{1-E^2}\; t + E a_1,
\label{combin}
\end{eqnarray}
this $\hat{t}$ has no dynamics. For $\widetilde{\mu}=0$, we obtain 
$E \p_1 \p_0 \hat{t}=0$ and this is
consistent with the fact that $\hat{t}$ has no dynamics.


For $\widetilde{\mu}= 2, \cdots, p$, the resultant equations of motion
are 
\begin{eqnarray}
\left(\p_0^2  -E^2 \p_1^2\right) a_\mu + E\p_1 \p_\mu \hat{t}=0.
\label{seni}
\end{eqnarray}
Therefore if we turn off the non-dynamical fluctuation $\hat{t}$, the
transverse fluctuation gauge fields $a_\mu$ are subject to the 2
dimensional Klein Gordon equation which shows the propagation with the
speed $E$. In other words, the causal structure of these fluctuations
$a_\mu$ $(\mu = 2, \cdots,p)$ is determined by the effective metric
(\ref{difca}). 

We have seen that a combination of $a_1$ and $t$ (\ref{combin}) is
non-dynamical. What about another combination which is orthogonal to
that? The equation for $\widetilde{\mu}=1$ is 
\begin{eqnarray}
  -E \p_0^2 t + \sqrt{1-E^2}\p_0^2 a_1 + E^2 \p_1^2 t =0.
\end{eqnarray}
Using (\ref{hatt}), the above equation is reduced to
\begin{eqnarray}
\left(\p_0^2  - E^2 \p_1^2\right) t =0.  
\end{eqnarray}
Therefore another combination also is subject to the effective metric
(\ref{difca}). 

After all, there are $p$ dynamical degrees of freedom which
propagates along $x^1$ with the speed $E$.

\subsection{Fluctuations and General BSFT type Action}

The analysis in section \ref{sena} took  full advantage of 
regarding the tachyon as a transverse scalar field. In this subsection, 
we show that the result above is quite universal, and does not refer to
any concrete form of the kinetic term function ${\cal F}(z)$ in
the Lagrangian. We utilize only the behavior of ${\cal F}(z)$ in the
rolling tachyon limit $z \rightarrow -1$. 

We use an alternative method of expanding the effective action itself. 
This is basically equivalent to analyzing the equations of motion 
as in section \ref{sena}. 
We expand the Lagrangian (\ref{geneact}), and collect all the terms
quadratic in the fluctuation. For our purpose it is clear that we don't
need the expansion of $V(T)$, since the expansion of the potential term
does not give derivatives of fluctuations and thus it is not related to
the causal structure which we want. 
After a straightforward calculation, we
obtain the expansion of ${\cal F}$ and the Born-Infeld term 
$\sqrt{-\det (\eta + F)}$ as
\begin{eqnarray}
 {\cal L} = -V(T_0)\sqrt{1-E^2}
\left[
 {\cal F}''(z^{(0)}) 
{\cal L}^{(2)} 
+ {\cal F}'(z^{(0)}) {\cal L}^{(1)}
+ {\cal F}(z^{(0)}) {\cal L}^{(0)}
\right],
\label{fl2a}
\end{eqnarray}
where $z^{(0)}$ is the background value of 
$z\equiv \widetilde{g}^{\mu\nu} \p_\mu T \p_\nu T$, 
\begin{eqnarray}
 z^{(0)} = \frac{-\dot{T}_0^2}{1-E^2},
\end{eqnarray}
and the components ${\cal L}^{(i)}$ ($n=0,1,2$) have the coefficient
$(\p_z)^n{\cal F}(z)|_{z=z^{(0)}}$ for each. 
The definition of them are as follows : 
\begin{eqnarray}
{\cal L}^{(2)}(t,a) 
& \equiv  &
\frac{2 \dot{T}_0^2}{(1-E^2)^2} \left(
\p_0 t + \frac{\dot{T}_0 E}{1-E^2}f_{01}
\right)^2, 
\label{l2}
\\
{\cal L}^{(1)}(t,a) &\equiv &
\left(
\frac{-1}{1-E^2}(\p_0 t)^2
+ \frac{1}{1-E^2}(\p_1 t)^2
+ (\p_i t)^2
\right) 
\nonumber
\\ 
& & +
\left(\frac{2E\dot{T}_0}{(1-E^2)^2} f_{10} \p_0 t
   +\frac{2E\dot{T}_0}{1-E^2} f_{i1} \p_i t \right)
\label{l1}
\\ 
& & 
+ 
\left(
\frac{-(1+E^2)\dot{T}_0^2}{(1-E^2)^3} (f_{01})^2
+\frac{-\dot{T}_0^2}{(1-E^2)^2} (f_{0i})^2
+\frac{E^2\dot{T}_0^2}{(1-E^2)^2} (f_{1i})^2
\right).
\nonumber
\end{eqnarray}
Here $i$ runs from $2$ to  $p$. 
The last component ${\cal L}^{(0)}$ includes $(f_{ij})^2$ for example,
however the expression for this term is unnecessary in the following
discussion.  

Now we observe that the original kinetic function ${\cal F}(z)$ has the 
following behavior in the rolling tachyon limit 
$z \rightarrow -1$ :  \begin{eqnarray}
 {\cal F}(z) \sim {\cal N}(1+z)^n \qquad (z \sim -1)
\end{eqnarray}
where ${\cal N}$ is a normalization constant, and 
a real number $n$ satisfies 
\begin{eqnarray}
 n<1, \quad n\neq 0.
\label{n}
\end{eqnarray}
This is derived from the fact that the
combination (\ref{combif}) is diverging in the limit.
The relation (\ref{n}) is satisfied for both Sen's action
($n=1/2$) and the BSFT action ($n=-1$). 
Then it follows that in the rolling tachyon limit 
\begin{eqnarray}
 |{\cal F}''(z^{(0)})| \gg 
 |{\cal F}'(z^{(0)})| \gg 
 |{\cal F}(z^{(0)})| .
\label{fff}
\end{eqnarray}
This means that in the expanded Lagrangian, the  dominant term is 
${\cal L}^{(2)}$ whose coefficient is most rapidly diverging. 
If we take $a_0=0$ gauge, then in the rolling tachyon limit 
$\dot{T}_0 \rightarrow \sqrt{1-E^2}$
the expression (\ref{l2}) is arranged as
\begin{eqnarray}
 {\cal L}^{(2)} = \frac{1}{(1-E^2)^{3/2}} (\p_0 \hat{t})^2
\label{ttef}
\end{eqnarray}
where we have used the linear combination (\ref{combin}).
This $\hat{t}$ is the mode parallel to the motion of the brane in the  
T-dual language which was studied in section \ref{sena}.
The Lagrangian for the mode $\hat{t}$ is effectively given by 
(\ref{ttef}) and other terms becomes irrelevant in the rolling tachyon
limit. Hence $\hat{t}$ becomes non-dynamical.

The dynamics for the gauge fields $a_i$ turns out to be different. 
The equation of motion for $a_i$ derived from the fluctuation Lagrangian
(\ref{fl2a}) is
\begin{eqnarray}
 \frac{2 \dot{T}_0^2}{(1-E^2)^2} \left(\p_0^2 - E^2 \p_1^2\right)a_i
 + \frac{2E}{1-E^2}\p_1 \p_i \hat{t} =0.
\end{eqnarray}
This is precisely what we have found in the analysis of 
Sen's action, (\ref{seni}).
Therefore if we turn off the non-dynamical field $\hat{t}$, then the
gauge field $a_i$ ($i=2,\cdots,p$) is subject to the effective metric 
(\ref{difca}). 

It is straightforward to see from the action (\ref{l1}), 
that another linear combination of $t$ and
$a_1$ which is orthogonal to $\hat{t}$ is subject again to the
effective metric (\ref{difca}) if we turn off the non-dynamical
$\hat{t}$. This coincides with the analysis in section \ref{sena}.

These results turn out to be quite universal, 
as follows from the simple limit behavior of the kinetic term of 
the action, (\ref{fff}). 

\section{Hamiltonian Dynamics and String-Like Objects}

Now that we have explored small fluctuations around homogeneous
vacua, let us turn to
dynamics of flux lines. The question of gauge dynamics upon 
tachyon condensation has an interesting history in recent years.
Tachyon condensation here is a process of unstable branes decaying,  
seen from open string degrees of freedom. Of various confusions
about this process, perhaps the most intriguing is what happens
to electric fluxes on the worldvolume of the decaying brane 
\cite{yi,conf,chicago}. 
Electric flux lines of the  Born-Infeld action carry fundamental
string charges from viewpoint of the spacetime, which is a
conserved charge.

While one might be content to say that the violent annihilation process
pushes all fluxes away to spatial infinity, just as most of worldvolume 
energy is dispersed, but this is not quite satisfactory. For 
instance, one can imagine setting up a device which selectively keeps 
a particular conserved charge. Something must be left behind and
carry that particular conserved charge, and we should then ask 
what object carries the leftover charge. The obvious answer to this,
in case of fundamental string charge  in the form of
electric flux lines, is of course the fundamental string itself
\cite{yi}. Less clear is precisely how this formation of
fundamental string from electric flux line happens. A nonperturbative
confinement mechanism has been proposed, while some advocate a
classical, still unknown, origin of such phenomenon.

Ref.~\cite{fluid} by 
two of the authors explored this question from purely classical
viewpoint. The exact dynamics of Sen's effective action contains a
fluid of flux lines at the vacuum, each of which behaves remarkably like 
an infinitesimal version of Nambu-Goto string.  
What is absent
in the classical dynamics is precisely something that will bind such
flux lines into a string-like objects of small width. In fact, a 
classically exact scale invariance was shown to exist and implies that
flux  
lines  have no transverse pressure whatsoever. On the other hand,
exactly because of this lack of pressure, the  possibility opens up that 
the slightest quantum correction could break the scale invariance and
drive the flux fluid into quantized units of flux bundles, which may 
then be identified with fundamental strings.

In any case, by an ansatz within the classical dynamics, 
one could still consider tightly bunched flux
lines, which thanks to the pressureless nature of the fluid, 
are of finite energy per length. When such a flux bundle is squeezed
into the form of delta function distribution, the resulting string-like
configuration follows the classical Nambu-Goto 
dynamics \cite{fluid,nielsen}.\footnote{This
feature turns out to extend to directions orthogonal to the
brane worldvolume, as was later demonstrated by A. Sen \cite{followup}.}  
In this
section, we would like to discuss how rolling tachyon background and 
the tachyon fluctuation figure into dynamics of flux lines.

\subsection{Small Fluctuations in Hamiltonian Viewpoint}

The derivation of Hamiltonian in Ref.~\cite{fluid} may be applied here
straightforwardly, since the tachyon kinetic terms enters as if it
were one of transverse scalars. The Hamiltonian of the combined
system of tachyon and gauge field is,
\begin{equation}
{\cal H}=\sqrt{ \pi^i\pi^i +P^2+(\pi^i\partial_i T)^2 + 
(F_{ij}\pi^j+\partial_i T\, P)^2 
+V^2\det(h)} ,
\end{equation}
where $h_{ij}\equiv \eta_{ij}+F_{ij} +\p_i T\p_j T $
with Roman indices $i,j$ running over all spatial directions,
$1,2,\dots,p$.
The vector $\pi_i$ is the conjugate momenta of the gauge field, which
corresponds to the conserved electric flux, or equivalently the
fundamental string density. Similarly $P$ is the conjugate momentum of
$T$. In the limit of tachyon condensation $V\rightarrow 0$, we find a
much simplified Hamiltonian,
\begin{equation}
{\cal H}=\sqrt{ \pi^i\pi^i +P^2+ (\pi^i\partial_i T)^2 +(F_{ij}\pi^j+
\partial_i T P)^2 } ,
\end{equation}
from which all dynamics, in principle,  can be read off.

There is a simpler way to express this Hamiltonian. We invent a new
direction and pretend that $T$ is a component of gauge field along 
this new imagined direction:
\begin{equation}
A_T=T
\end{equation}
and similarly $\pi^T=P$. Then the Hamiltonian is succinctly written as
\begin{equation}
{\cal H}=\sqrt{ \pi^M\pi^M +(F_{MK}\pi^K)^2 } ,
\end{equation}
with the understanding that $\p_T\equiv 0$ on any object. Indices $M,K$
runs over $1,2, \dots, p$ and $ T$.
Hamiltonian equations of motion are then,
\begin{eqnarray}
\dot A_M &=&\frac{1}{\cal H} \left( \pi_M + F_{LK}\pi^K F_{LM} \right), 
\nonumber \\
\dot \pi^M &=& \p_L \left(\frac{1}{\cal H} \left(\pi^MF_{LK}\pi^K -
 \pi^L F_{MK}\pi^K \right)\right) .
\end{eqnarray}
Let us expand 
\begin{eqnarray}
\pi^M&=& (1+\epsilon)\pi^M_0+\delta\pi^M ,\nonumber \\
F_{MK}&=&\delta F_{MK} ,
\end{eqnarray}
around a constant and uniform $\pi_0^M$ background, which could consist
of any combination of $\dot T $ and $E$. We define $\delta \pi^M$ to be 
orthogonal to the background value $\pi_0^M$, and encode the rest
in $\epsilon$.
Taking antisymmetric derivative of $\dot A_M$ equation, and truncating
to the first order perturbation, we find
\begin{equation}
\delta \dot F_{MK}=\frac{1}{\cal H}_0
\left(\p_M\delta \pi_K-\p_K\delta \pi_M\right).
\end{equation}
Taking time derivative of the $\pi$ equation and inserting this
relationship, 
\begin{equation}
\ddot \epsilon = -\frac{1}{{\cal H}_0^2}
(\pi^K_0 \p_K)(\p_L \delta \pi^L),
\end{equation}
which shows that $\epsilon\pi_0^M$, one linear combinations of $A_i$ and
$T$,  is not dynamical. Rather its evolution is determined completely by
that of  $\delta \pi^M$.  This may be regarded as a consequence of
Gauss's constraint, even though it involves fluctuation of $T$ which is
not charged. The orthogonal part, on the other hand, obeys a dynamical
equation 
\begin{equation}
\delta \ddot \pi^M= \frac{1}{{\cal H}_0^2}(\pi^K_0 \p_K)^2\delta \pi^M ,
\end{equation}
which are 1+1 dimensional Klein-Gordon equations.

Recall that $\partial_T\equiv 0$ by definition. Labelling the
direction of $\vec E$ (thus that of $\vec\pi$) as 1, we have
\begin{equation}
\left(\p_0^2 - E^2\p_1^2\right)\delta \pi^M =0 ,
\end{equation}
the most general solution to which has the form
\begin{equation}
h_L(Ex^0+x^1,x^2,x^3,\dots,x^p)+h_R(Ex^0-x^1,x^2,x^3,\dots,x^p).
\end{equation}
This shows that, although $\delta \pi^M$ 
may have dependence on coordinates
transverse to $\vec E$, such dependence does not propagate.
Rather only variation along $\vec E \cdot \vec x$ may 
propagate with speed $E$. This faithfully reproduces behavior found in
previous section via analysis of the effective metric. For the purely
rolling tachyon limit of $E=0$, the solution has no time-dependence
at all, matching exactly onto the Carrollian behavior.

Thus, in general, $p$ degrees of freedom propagate along $\vec E=\vec 
\pi/{\cal H}$ at speed $|\vec E|$.
This analysis can be repeated in the presence of transverse scalars
$X^I$ in exactly the same way. Again we may treat transverse scalars
$X^I$, $I=p+1,p+2,\dots, D-1$, as if they were gauge fields along
additional directions. For an unstable brane in $D$ spacetime
dimensions, we find   $D-1$  degrees of freedom
propagating unidirectionally  along the vector $\pm\vec E$ at speed
$|\vec E|$. 

\subsection{String Fluid in Rolling Tachyon Background}

There are a few difficulties one encounter in constructing string-like
objects in this classical setting. For instance, let us consider the
following class of solutions to the Hamiltonian equation of motion;
\begin{eqnarray}
&&\pi^M=\hat m^M \Phi(x) , \nonumber\\
&&F_{MK}\hat m^K=0 ,
\end{eqnarray}
with a constant unit vector $\hat m$, provided that
\begin{equation}
\hat m^M\partial_M \Phi(x) =0.
\end{equation}
In terms of $P$ and $\vec \pi$, this means that we have
a family of solutions where,
\begin{eqnarray}
&&\pi^i=\hat n^i \cos\theta \,\Phi(x) ,\nonumber\\
&&P=  \sin\theta \;\Phi(x) ,
\end{eqnarray}
with 
\begin{equation}
\hat n^i\p_i\Phi=0 ,
\end{equation}
and also 
\begin{eqnarray}
&&\hat n^i\p_i T=0 , \nonumber \\
&&\sin\theta\,\p_i T= \cos\theta\,\hat n^j F_{ji} ,
\end{eqnarray}
with an arbitrary angle $\theta$. Unit vector $\hat n$ is related to
$\hat m$ above by
\begin{equation}
\hat m^M =(\cos\theta\,\hat n^i, \sin\theta ).
\end{equation}
It is straightforward to verify that such a configuration solves the
Hamiltonian equation of motion trivially.

This class of solutions, where both flux density $|\vec \pi|$
and the Hamiltonian $\cal H$ is proportional to $\Phi$, 
represents an arbitrary transverse distribution of flux lines 
accompanied by tachyon matter following the same pattern of 
distribution.
\begin{eqnarray}
|\vec \pi| &=& \cos\theta\,\Phi , \nonumber \\
P &=& \sin\theta\,\Phi ,\nonumber \\
{\cal H}&=& \Phi .
\end{eqnarray}
Typically there are  infinitely many  possible choices of $\Phi$ allowed
by 
the constraints, so the same sort of infinite degeneracy that had 
plagued previous attempts at constructing tightly bound string-like 
objects, shows up here again. There are infinite number of
possibilities for $\Phi$ with a fixed and finite total flux
and thus with the same total energy.

With $P$ approaching some constant value at infinity, we must
be more careful for we wish to consider finite total flux. 
The above class of solution indicates that flux distribution
tends to match that of $P$ and spread out all over the space,
but a more detailed analysis is required.
Regardless of any such detail, however, this modification
cannot help for the 
following reason: In the  actual tachyon condensation of unstable
branes, the decaying brane is coupled to closed strings
and will lose energy by emission of the latter degrees of 
freedom. If we wish to look for fundamental string in the
final vacuum state, we should set $P=0={\cal H}$ in the
asymptotic region. This brings us back to  the degeneracy problem above.

\subsection{Closed Flux String}

With possibility of rolling tachyon mixing with $\vec E$, on the other 
hand, we are able to remove one other difficulty in constructing
closed-string-like
object. Without the rolling tachyon, we have nowhere vanishing electric
field with $|\vec E|=1$,
which makes it difficult to envisage a closed loop of electric flux. 
One ends
up with singular configuration of $\vec E$ necessarily since closed
loop of flux lines involve a winding number 
\begin{equation}
\oint d\vec l \cdot \vec E ,
\end{equation}
which cannot be smoothly taken away unless $E=0$ somewhere. While
the singularity does not translate to divergent energy or action
in this case, it is still an uncomfortable situation.

When the rolling tachyon is taken into account, the prospect looks
better.  
Since the vacuum condition is modified to $E^2+\dot T^2=1$,
$E$ can take any value between 0 and 1 and closed loops of flux 
lines are much easier to conjure. For instance, we may imagine a tightly  
bunched flux lines, outside of which the vacuum is dominated by rolling 
tachyon $\dot T=1$. Since $|\vec E|$ is no longer restricted to be 1, 
the potential singularity due to the winding number is naturally
resolved.  

One very intriguing aspect of such configurations is that
the  propagation
of small fluctuation is allowed only near the core where flux lines
are densely populated. For instance, we could consider a very tight
flux bundle, at the centre  of which $|\vec E|$ approaches $1$. This
can be easily achieved by having a divergent $\vec \pi$ at the core.
The only propagation of signals allowed are then along the 1+1 
dimensional core of the bundle, where signals propagate at speed of 
light. Outside, where $E=0$, no small fluctuation is allowed since
the causal structure became Carrollian.

This way, we have conjured up a 1+1 dimensional coherent object
initially realized on $p+1$ dimensional worldvolume. Furthermore,
there are $D-1$ degrees of freedom propagating only along the core 
of the object. The configuration could be closed or infinitely 
extended. This looks remarkably similar to a relativistic string,
except for one problem; it appears that classical dynamics has
no mechanism to bind flux lines into such a tight bundle. This can be
also seen from the fact that the number of propagating degrees of
freedom is  
$D-1$, instead of $D-2$ we came to expect from a relativistic Nambu-Goto 
string; the additional degree of freedom dictates how the flux lines are
distributed over the transverse directions. We believe that there is
a quantum mechanical reason for binding of flux lines, without which
proper understanding of closed string is not possible \cite{yi,conf}.

\section{Conclusion and Discussions} 

In this paper, we studied classical systems consisting of open string
tachyon 
coupled to gauge field. Since the dissipative couplings to
closed string degrees of freedom are suppressed, the system is conserved 
and the initial energy of D-branes and their worldvolume fields is
retained in the form of tachyon matter and fluid of electric flux
lines. Tachyon condensation is specified by a combination of rolling
tachyon $\dot T$ and background electric field $\vec E$, satisfying
$E^2+\dot T^2=1$. The purely rolling tachyon background, $\dot T=1$,
forces a Carrollian limit for all small fluctuation, and this explains
the absence of perturbative degrees of freedom. On the other hand,
general 
vacuum with $|\vec E|\neq 0$ allows 1+1 dimensional propagation of small
fluctuations.  In this case, the signal propagate at speed $|\vec E|$.

In section 4, we have seen that the causal structure of the gauge and
tachyon fluctuations in the rolling tachyon phase is universal in that
it does not depend much on the structure of the kinetic
function ${\cal F}(z)$. What is required for the form of the function 
${\cal F}(z)$ is only to have a rolling tachyon solution, 
$E^2 + \dot{T}^2 =1$. In fact, the effective actions of 
non-BPS branes can be determined up to the field redefinition, which
corresponds to the choice of the renormalization scheme in deriving the
effective actions. For example, one may change 
$T \rightarrow T + a(T) z + \cdots$, as results in the change of the
function ${\cal F}(z)$ \cite{Muk}. 
In order to study the dynamics without 
referring to specific forms of the effective action, one may come back
to the BSFT formulation and consider the rolling tachyon background
there. In the appendix, we make a brief comment on this for a special
case $E=0$. A more general study will be necessary for studying the
causal structure of the massive modes and general open string
excitations in the rolling tachyon phase. 

Although in this paper and most of the literature an 
homogeneous tachyon profile has been employed as an initial condition, 
it is expected that in general the tachyon may roll down
inhomogeneously \cite{inhomo}.  It was shown in Ref.\ \cite{time} 
that inhomogeneous decay of non-BPS brane may hit a singularity of
localized energy in the
rolling process and this can be interpreted as the  formation of lower
dimensional branes. If we include a background electric field in
that setup, it might describe a formation of 
(F, D) bound states. In this respect it would be intriguing to analyse
the effective metric and fluctuation dynamics in the inhomogeneous
rolling tachyon background with electric fields. 

In particular, one of more important outstanding questions is how
inhomogeneous distribution of electric flux lines evolves in time. As we
saw in section 5, there appears to be no reason to believe that fluxes
would come together and form a localized string-like object. On the
other hand, the causal structure found above has interesting
implications once the flux lines are somehow grouped into well-separated
bundles of flux strings; it tells us that the only degrees of freedom
remaining would be those that propagate along the flux string, at speed
of the light, and dictate how the flux string moves. Such a flux string
will behave like a classical Nambu-Goto string \cite{fluid}. While the
classical dynamics above has one additional degree of freedom which
must be related to the thickness of such a flux string, this must be lifted
somehow by the (still unknown) binding mechanism. 

\vskip 1cm

\noindent
\leftline{\Large \bf Acknowledgments}
\vskip 0.5cm \noindent
A part of this work was done during the KIAS workshop on Strings
and Branes.  G.\ G.\ and K.\ H.\ are grateful to KIAS for kind
hospitality. K.\ H.\ would like to express his gratitude to 
String Theory Group in National Taiwan University for helpful 
support, and would like to thank O.\ Andreev for useful
comments. P.\ Y.\ would like to thank Theory Group of Columbia
University for its hospitality, where part of the manuscript is
written.

\vskip 1cm
\appendix

\section{Universality of the open string metric}

To generalize our 
result to the one valid for other open string excitations, 
it is useful to remember  how the effective
open string metric $G^{\rm open}_{\mu\nu}$ was derived in string
theory. If the gauge field strength acquires a vev, the effective metric
is obtained by the two point function on the boundary of the disc
worldsheet with 
infinite number of the boundary insertion of the condensed gauge
field. This effectively changes the boundary condition of a worldsheet
and the Green function. The tensor coefficient of this 
Green function $\langle X^\mu X^\nu\rangle$ gives the effective metric.  

Therefore, we consider the two point function of a special 
two dimensional theory with the boundary insertion corresponding to the
rolling tachyon. In the BSFT formalism the kinetic term of the open
string component field  $f^{\mu_!, \cdots, \mu_n}$ is given by the two
point function 
\begin{eqnarray}
 \langle V(f) V(f) \rangle,
\end{eqnarray}
which is evaluated using the following 
two dimensional field theory \cite{BSFT}: 
\begin{eqnarray}
 L = \frac{1}{8\pi}
\int_\Sigma d^2 \sigma \sqrt{h} h^{\alpha\beta} \partial_\alpha X
\partial_\beta X + \frac{u}{8\pi}
\oint_{\partial \Sigma} d\theta X^2.
\label{acbs}
\end{eqnarray}
In the two point function, the indices of the field $f$ are related to
the $X^{\mu_i}$ operator in the vertex operators $V(f)$. The boundary 
Green function in this two dimensional field theory after the
regularization was given in Ref.\ \cite{BSFT} as 
\begin{eqnarray}
 G(\theta, \theta')= 2 \sum_{k\in Z} \frac{1}{|k|+u} 
\exp (ik(\theta-\theta')). 
\label{propa}
\end{eqnarray}
Here note that $u$ should be positive in the original BSFT where the
boundary interaction is for the spatial $X$, but in the rolling tachyon 
we have to consider this $X$ as $X^0$ and thus $u$ is 
effectively negative, with respect to the first term in the action
(\ref{acbs}). If $u$ is 
positive, the above Green function is finite and well defined. However, 
if $u$ is negative, and especially for $u=-1$, the above Green function
diverges. In fact, for the rolling tachyon limit $u=-(\dot{T})^2 = -1$.
Therefore, the time component of the co-metric is divergent, hence
the effective metric appearing in the kinetic terms of all the open 
string excitations becomes Carrollian in the rolling tachyon limit.
This is consistent with the results in this paper, since as seen 
in section 4, especially when $E=0$ all
the gauge fluctuations are governed also by the Carrollian metric.
With more precise evaluation of the propagator (\ref{propa}), we hope
that the exact form of the effective metric (\ref{inverse}) may be
obtained.  

This argument is valid only in bosonic string theory. In 
superstring theory, one has  two kinds of two point functions :  
$\langle X^{\mu} X^{\nu} \rangle$ and 
$\langle\psi^{\mu} \psi^{\nu} \rangle$. Thus the metric appearing in the
effective action is generically a combination of these. This combination
can be different for every excited state, and actually 
that for the tachyon and the gauge field are different 
\cite{Andreev}.


\newcommand{\J}[4]{{\sl #1} {\bf #2} (#3) #4}
\newcommand{\andJ}[3]{{\bf #1} (#2) #3}
\newcommand{\AP}{Ann.\ Phys.\ (N.Y.)}
\newcommand{\MPL}{Mod.\ Phys.\ Lett.}
\newcommand{\NP}{Nucl.\ Phys.}
\newcommand{\PL}{Phys.\ Lett.}
\newcommand{\PR}{ Phys.\ Rev.}
\newcommand{\PRL}{Phys.\ Rev.\ Lett.}
\newcommand{\PTP}{Prog.\ Theor.\ Phys.}
\newcommand{\hep}[1]{{\tt hep-th/{#1}}}


\begin{thebibliography}{99}


\bibitem{DaD}
A.~Sen,
``Tachyon condensation on the brane anti-brane system,''
\J{JHEP}{08}{1998}{012}, {\tt hep-th/9805170}.

\bibitem{emil}     
J.~A.~Harvey, D.~Kutasov and E.~J.~Martinec,
``On the relevance of tachyons,'' {\tt hep-th/0003101}.

\bibitem{BSFT}
E.\ Witten,
  { ``On Background Independent Open String Field Theory,''}
  \J{\PR}{D46}{1992}{5467}, {\tt hep-th/9208027}; 
  { ``Some Computations in Background Independent Off-Shell String
    Theory,''}
  \J{\PR}{D47}{1993}{3405}, {\tt hep-th/9210065}.

\bibitem{BSFTo}
  K.\ Li and E.\ Witten, 
  { ``Role of Short Distance Behavior in Off-Shell Open String
    Field Theory,''} 
  \J{\PR}{D48}{1993}{853}, {\tt hep-th/9303067};

  S.\ Shatashvili,
  { ``Comment on The Background Independent Open String Theory,''}
  \J{\PL}{B311}{1993}{83}, {\tt hep-th/9303143}; 
  { ``On The Problems with Background Independence in String
    Theory,''} {\tt hep-th/9311177}.



\bibitem{GS}
  A.\ A.\ Gerasimov and S.\ L.\ Shatashvili, 
  { ``On Exact Tachyon Potential in Open String Field Theory,''}
  \J{JHEP}{0010}{2000}{034}, {\tt hep-th/0009103}.

  D. Kutasov, M.\ Marino and G.\ Moore,
  { ``Some Exact Results on Tachyon Condensation in String Field
    Theory,''}
   \J{JHEP}{0010}{2000}{045}, {\tt hep-th/0009148}.

  D.\ Ghoshal and A.\ Sen, 
    { ``Normalization of the Background Independent Open String Field
             Theory Action''}, \J{JHEP}{0011}{2000}{021},  
   {\tt hep-th/0009191}

  M.\ Marino,     { ``On the BV formulation of boundary superstring
    field theory,''} \J{JHEP}{0106}{2001}{059},
   {\tt hep-th/0103089}.

  V.\ Niarchos and N.\ Prezas,
    { ``Boundary Superstring Field Theory,''}
  \J{\NP}{B619}{2001}{51}, {\tt hep-th/0103102}.

\bibitem{kutasov}
D.\ Kutasov, M.\ Marino and G.\ Moore,
``Remarks on Tachyon Condensation in Superstring Field Theory,''
 {\tt hep-th/0010108}.

\bibitem{rolling}
A.~Sen,
``Rolling tachyon,''
\J{JHEP}{0204}{2002}{048},
{\tt hep-th/0203211};
``Tachyon matter,''
{\tt hep-th/0203265};
``Field theory of tachyon matter,''
{\tt hep-th/0204143}.

\bibitem{bergshoeff}
M.\ R. Garousi, 
``Tachyon couplings on non-BPS D-branes and Dirac-Born-Infeld action,'' 
\J{\NP}{B584}{2000}{284}, {\tt hep-th/0003122}.

E.\ A.\ Bergshoeff, M.\ de Roo, T.\ C.\ de Wit, E.\ Eyras and S.\ Panda, 
``T-duality and Actions for Non-BPS D-branes,''
\J{JHEP}{0005}{2000}{009}, {\tt hep-th/0003221}.

\bibitem{kluson}
J.~Kluson,
``Proposal for non-BPS D-brane action,''
\J{\PR}{D62}{2000}{126003}, {\tt hep-th/0004106};
``D-branes from N non-BPS D0-branes,''
\J{JHEP}{0011}{2000}{016}, {\tt hep-th/0009189};

J.~A.~Minahan and B.~Zwiebach,
``Gauge fields and fermions in tachyon effective field theories,''
\J{JHEP}{0102}{034}{2001}, {\tt hep-th/0011226};

M.~Alishahiha, H.~Ita and Y.~Oz,
``On superconnections and the tachyon effective action,''
\J{\PL}{B503}{2001}{181}, {\tt hep-th/0012222}.


\bibitem{fluid}
G.~W.~Gibbons, K.~Hori and P.~Yi,
``String fluid from unstable D-branes,''
\J{\NP}{B596}{2001}{136}, {\tt hep-th/0009061}.

\bibitem{potential}
A.~Sen,
``Supersymmetric world-volume action for non-BPS D-brane,''
\J{JHEP}{9910}{1999}{008}, {\tt hep-th/9909062};
``Universality of the tachyon potential,'' 
\J{JHEP}{9912}{1999}{027}, {\tt hep-th/9911116}.

\bibitem{followup}
A.~Sen,
``Fundamental strings in open string theory at the tachyonic vacuum,''
\J{J.\ Math.\ Phys.}{42}{2001}{2844}, {\tt hep-th/0010240}.

\bibitem{yi}
P.~Yi,
``Membranes from five-branes and fundamental strings from Dp branes,''
\J{\NP}{B550}{1999}{214}, {\tt hep-th/9901159}.

\bibitem{conf}
O.~Bergman, K.~Hori and P.~Yi,
``Confinement on the brane,''
\J{\NP}{B580}{2000}{289},
{\tt hep-th/0002223}.

\bibitem{shenker}
M. Kleban, A. Lawrence and S. Shenker,
``Closed strings from nothing,''
\J{\PR}{D64}{2001}{066002}, {\tt hep-th/0012081}.

\bibitem{Leblond} J.\ M.\ L\'evy-Leblond, ``Une nouvelle limite
non-relativiste du group de Poincar\'e,'' \J{Ann.\ Inst.\ H.\
Poincar\'e}{3}{1965}{1}.

\bibitem{Sen} N.\ D.\ Sen Gupta, ``On an Analogue of the Galileo
Group,'' \J{Il Nuovo Cimento}{44A}{1966}{512}.

\bibitem{Bacry} H.\ Bacry and J.\ M.\ L\'evy-Leblond, ``Possible
Kinematics,'' \J{J.\ Math.\ Phys.}{9}{1967}{1605}.

\bibitem{Nuyts} H.\ Bacry and J.\ Nuyts,
 ``Classification of Ten-Dimensional Kinematical Groups with Space
 Isotropy,'' \J{J.\ Math.\ Phys.}{27}{1986}{2455}.
 
\bibitem{Gibbons1} G.\ W.\ Gibbons,   ``Cosmological Evolution 
of the Rolling Tachyon,'' \J{\PL}{B537}{2002}{1}, 
{\tt hep-th/0204008}.

\bibitem{cos}
M.~Fairbairn and M.~H.~Tytgat, 
{\tt hep-th/0204070};
S.~Mukohyama, 
\J{\PR}{D66}{2002}{024009},
{\tt hep-th/0204084};
A.~Feinstein,
{\tt hep-th/0204140};
T.~Padmanabhan,
\J{\PR}{D66}{2002}{021301},
{\tt hep-th/0204150};
A.~Frolov, L.~Kofman and A.~A.~Starobinsky, 
{\tt hep-th/0204187};
D.~Choudhury, D.~Ghoshal, D.~P.~Jatkar and S.~Panda, 
{\tt hep-th/0204204};
X.~z.~Li, J.~g.~Hao and D.~j.~Liu, 
{\tt hep-th/0204252};
G.~Shiu and I.~Wasserman, 
\J{\PL}{B541}{2002}{6}, 
{\tt hep-th/0205003};
T.~Padmanabhan and T.~R.~Choudhury, 
the  same scalar field?,'' 
{\tt hep-th/0205055};
L.~Kofman and A.~Linde, 
\J{JHEP}{0207}{2002}{004}, 
{\tt hep-th/0205121};
H.~B.~Benaoum, 
{\tt hep-th/0205140};
M.~Sami, 
{\tt hep-th/0205146};
M.~Sami, P.~Chingangbam and T.~Qureshi, 
{\tt hep-th/0205179};
T.~Mehen and B.~Wecht, 
{\tt hep-th/0206212};
G.~Shiu, S.~H.~Tye and I.~Wasserman, 
{\tt hep-th/0207119};
Y.~S.~Piao, R.~G.~Cai, X.~m.~Zhang and Y.~Z.~Zhang, 
{\tt hep-ph/0207143};
X.~z.~Li, D.~j.~Liu and J.~g.~Hao,
{\tt hep-th/0207146};
J.~M.~Cline, H.~Firouzjahi and P.~Martineau,
{\tt hep-th/0207156};
B.~Wang, E.~Abdalla and R.~K.~Su,
{\tt hep-th/0208023};
S.~Mukohyama,
{\tt hep-th/0208094};
M.~C.~Bento, O.~Bertolami and A.~A.~Sen,
{\tt hep-th/0208124}.

\bibitem{Herdeiro} G.\ W.\ Gibbons and C.\ A.\ R.\
Herdeiro, ``Born-Infeld Theory and Stringy Cosmology,'' 
\J{\PR}{D63}{2001}{064006},  {\tt hep-th/0008052}.

\bibitem{Gibbons2} G.\ W.\ Gibbons, 
``Aspects of Born-Infeld Theory and String/M-Theory,'' \\
{\tt hep-th/0106059}.

\bibitem{Gibbons3} G.\ W.\ Gibbons, ``Pulse Propagation in Born-Infeld
theory, the World Volume Equivalence Principle and the Hagedron-like
Equation of State of the Chaplygin Gas,'' 
\J{Grav.\ Cosmol.}{8}{2002}{2},  {\tt hep-th/0104015}.

\bibitem{West} G.\ W.\ Gibbons and P.\ C.\ West, 
``The metric and strong coupling limit of the M5-brane,'' 
\J{J.\ Math.\ Phys.}{42}{2001}{3188},  {\tt hep-th/0011149}.


\bibitem{Wald} S.\ Gao and R.\ M.\ Wald, 
``Theorems on  gravitational time delay and related issues,'' 
\J{Class.\ Quant.\ Grav.}{17}{2000}{4999},  {\tt gr-qc/0007021}.

\bibitem{Page} D.\ N.\ Page, S.\ Surya and E.\ Woolgar, 
``Positive Mass from Holographic Causality,'' 
{\tt hep-th/0204198}.

\bibitem{Hawking} S.\ W.\ Hawking, ``The Conservation of Matter in
General Relativity,'' \J{Comm.\ Math.\ Phys.}{18}{1970}{303}.

\bibitem{Carter} B.\ Carter, ``Energy Dominance and the Hawking
Ellis vacuum conservation theorem,'' {\tt gr-qc/0205010}.

\bibitem{Terashima}
S.\ Sugimoto and S.\ Terashima, 
``Tachyon Matter in Boundary String Field Theory,''
\J{JHEP}{0207}{2002}{025},
{\tt hep-th/0205085}.

J.\ A.\ Minahan, 
``Rolling the tachyon in super BSFT,''
\J{JHEP}{0207}{2002}{030}, {\tt hep-th/0205098}.


\bibitem{Ishida}
A.\ Ishida and S.\ Uehara,
``Gauge Fields on Tachyon Matter,''  {\tt hep-th/0206102}. 

T.\ Mehen and B.\ Wecht, 
``Gauge Fields and Scalars in Rolling Tachyon Backgrounds,''
{\tt hep-th/0206212}.

\bibitem{Muk}
P.~Mukhopadhyay and A.~Sen,
``Decay of Unstable D-branes with Electric Field,''
{\tt hep-th/0208142}.

\bibitem{LS}
N.\ D.\ Lambert and I.\ Sachs, 
``On Higher Derivative Terms in Tachyon Effective Actions,''
\J{JHEP}{0106}{2001}{060}, {\tt hep-th/0104218};
``Tachyon Dynamics and Effective Action Approximation,''
{\tt hep-th/0208217}.

\bibitem{supertube}
D.~Mateos and P.~K.~Townsend,
``Supertubes,''
\J{\PRL}{87}{2001}{011602}, {\tt hep-th/0103030}.

\bibitem{BIparticle}
G.~W.~Gibbons,
``Born-Infeld particles and Dirichlet p-branes,''
\J{\NP}{B514}{1998}{603}, {\tt hep-th/9709027}.

\bibitem{Hashimoto}
K.~Hashimoto, T.~Hirayama and S.~Moriyama,
``Symmetry origin of nonlinear monopole,''
\J{JHEP}{0011}{2000}{014}, {\tt hep-th/0010026}.

\bibitem{Bachas}
C.\ Bachas 
``D-brane dynamics,''
\J{\PL}{B374}{1996}{37}, {\tt hep-th/9511043}.

%

\bibitem{chicago}
J.\ A.\ Harvey, P.\ Kraus, F.\ Larsen and E.\ J.\ Martinec,
``D-branes and Strings as Non-commutative Solitons,''
\J{JHEP}{0007}{2000}{042}, {\tt hep-th/0005031}.

\bibitem{nielsen}
H.\ B.\ Nielsen and P.\ Olesen, 
``Local field theory of the dual string,''
\J{\NP}{B57}{1973}{367}.

\bibitem{inhomo}
B.~Craps, P.~Kraus and F.~Larsen,
``Loop corrected tachyon condensation,''
\J{JHEP}{0106}{2001}{062}, {\tt hep-th/0105227};

K.~Hashimoto,
``Dynamical decay of brane-antibrane and dielectric brane,''
\J{JHEP}{0207}{2002}{035}, {\tt hep-th/0204203};

G.~N.~Felder, L.~Kofman and A.~Starobinsky,
``Caustics in tachyon matter and other Born-Infeld scalars,''
{\tt hep-th/0208019}.


\bibitem{time}
A.~Sen,
``Time evolution in open string theory,''
{\tt hep-th/0207105}.

\bibitem{Andreev}
O.\ Andreev, 
``Some Computations of Partition Functions and Tachyon Potentials in
        Background Independent Off-Shell String Theory,''
\J{\NP}{B598}{2001}{151}, {\tt hep-th/0010218}.

\end{thebibliography}
\end{document}